\documentclass
{mn2e}
\usepackage{epsfig}
\usepackage{amsmath, amssymb,bm}

\def \be{\begin{equation}}
\def \ee{\end{equation}}

\def \msun{\rm M_{\odot}}

\def  \le{{L_{\rm Edd}}}

\begin{document}
\title[The Black--Hole Masses of High--Redshift QSOs] {The Black--Hole Masses of High--Redshift QSOs}

\author[Andrew King] 
{\parbox{5in}{Andrew King$^{1, 2, 3}$ 
}
\vspace{0.1in} \\ $^1$ School of Physics \& Astronomy, University
of Leicester, Leicester LE1 7RH UK\\ 
$^2$ Astronomical Institute Anton Pannekoek, University of Amsterdam, Science Park 904, NL-1098 XH Amsterdam, The Netherlands \\
$^{3}$ Leiden Observatory, Leiden University, Niels Bohrweg 2, NL-2333 CA Leiden, The Netherlands}

\maketitle

\begin{abstract}
Observations of high--redshift quasars frequently promote suggestions
of large black hole masses, whose presence so early in cosmic time is not easily explicable. I consider the parallel with 
ultraluminous X--ray sources (ULXs) --
now known to be
stellar--mass black hole (and neutron star) binaries apparently radiating far above their Eddington luminosities $L_{\rm Edd}$. The true
luminosity in ULXs is actually only of order $L_{\rm Edd}$, 
for {\it stellar--mass} accretors, 
but has a very anisotropic (`beamed') component, plus a near--isotropic component of similar luminosity but much lower specific intensity. Observers viewing ULXs from within the beam but assuming spherical symmetry deduce a luminosity $\gg L_{\rm Edd}$. These features appear because the accretors are fed mass at highly super--Eddington rates, most of it expelled in high--speed ($v >0.2c$) outflows from the accretion disc. 

I show that in similarly--beamed AGN,
emission--line properties would be essentially the same as in unbeamed sources, but
standard virial mass indicators
unusable because velocity widths are dominated by the outflows, not bound motions about the black holes. In an ensemble of this kind
the apparently most luminous systems are always the most distant, but have the lowest black hole masses. Interpreting observations of this ensemble without knowing that they are beamed leads instead to very high black hole mass estimates. The analogy with ULXs therefore suggests that high--redshift quasars might actually have central black hole masses which could have grown from stellar values within the lookback time.
I consider how one might test these ideas observationally.
\end{abstract}


\footnotetext[1]{E-mail: ark@astro.le.ac.uk}
\section{Introduction}
\label{intro}

Observations of quasars at redshift $\sim 6$ are often interpreted as requiring black hole masses $M \gtrsim 3\times 10^9\msun$ (see e.g. Willott, McLure \& Jarvis, 2003 for a very clear description). Depending on the detailed accretion history, these lead to suggestions that these holes descend from very large black hole seed masses $\gtrsim 10^8\msun$ which must have formed very rapidly at very high redshift, and there have been many efforts to model what these might be (e.g. Begelman, Volonteri \& Rees, 2006; Begelman, 2009; Regan \& Haehnelt, 2013; Johnson et al., 2013)\footnote{One can avoid the need for massive seeds if the accretion process keeps the black hole spin low. This in turn keeps the accretion efficiency low, so that the same luminosity corresponds to a higher mass. This can happen if accretion on to the hole is chaotic, i.e. from a succession of accretion events with uncorrelated angular momenta. See King, Pringle \& Hofmann, 2008 and Section 5.4 of King (2023)   for a detailed discussion of this point.}.

An important part of the argument leading to this is the assumption that the apparent luminosity\footnote{In this paper I will use the term `apparent luminosity' and the symbol $L_{\rm sph}$ to denote cases where emission is assumed isotropic, even though it may in reality be strongly anisotropic.}
$L \sim \nu L_{\nu}$ (where $L_{\nu}$ is the monochromatic luminosity, assumed isotropic) of the observed quasars cannot significantly exceed the Eddington limit 
\be
\le \simeq 1.3\times 10^{46}M_8\, {\rm erg\, s^{-1}}
\label{ledd}
\ee
for black hole mass $M= 10^8\msun M_8$.

Accretion on to a black hole is an essentially scale--free process, since the mass $M$ of the accretor defines typical lengthscales through the gravitational radius $R_g = GM/c^2$, timescales through $R_g/c$, and luminosities through $\le(M)$. 
Accordingly it is often helpful to compare analogous
accreting stellar--mass and supermassive black hole systems. Here I ask what we can learn from the stellar--mass version of the argument leading to very large black hole masses for high--redshift quasars.

\section{Ultraluminous X--ray Sources}

The stellar--mass analogues of quasars with very high deduced black hole masses are the ultraluminous X--ray sources (ULXs). About 1800 systems are known by now (see King, Lasota \& Middleton, 2023 for a recent review). 

ULXs are extragalactic systems which are not located in the centres of their host galaxies, but whose apparent luminosities ($\gg 10^{39}\, {\rm erg\, s}^{-1}$), if assumed isotropic,  require (through (\ref{ledd})) accretor masses significantly larger 
than likely for black holes descending from stellar evolution. Colbert \& Mushotzky (1999) identified ULXs as a class,
and suggested that they were systems with accretion on to hypothetical `intermediate--mass' black holes (or IMBH), where the hole mass was significantly larger ($\sim 10^4 - 10^5\msun$) than could result from stellar evolution. 

But King et al. (2001) noted that mass transfer at super--Eddington rates in otherwise standard X--ray binaries (with normal stellar--mass accretors) was very likely to lead to strong ejection of dense winds from the accretion disc, as pointed out by Shakura \& Sunyaev (1973). That paper had shown that a disc accreting at a rate which exceeds the local radiation pressure limit within a certain disc radius would eject the excess at each radius further in, ultimately producing total luminosity 
\be
L = L_{\rm Edd}(1 + \ln\dot m),
\ee
where $\dot m = \dot M_{\rm tr}/\dot M_{\rm Edd} \gg 1$ is the ratio of the rate
$\dot M_{\rm tr}$ of mass transfer towards the accretor
to the value $\dot M_{\rm Edd} = L_{\rm Edd}/\eta c^2$ (with $\eta \simeq 0.1$ the accretion efficiency, given by the specific binding energy of the black hole ISCO radius), which is the rate at which the accretor mass actually grows.

King et al. (2001) suggested that this would make a significant part of the luminosity
strongly anisotropic, i.e. beamed\footnote{In this paper `beaming' simply means collimated by scattering, and not more exotic processes such as relativistic beaming.}, as much of the accretion luminosity emitted from close to the black hole finds the two narrow vacuum funnels along the rotational axis of the accretion disc and escapes along them. The disc emission which does not find these funnels produces a near--isotropic luminosity of a similar order $\sim L_{\rm Edd}$. This is emitted from the outer photosphere of the wind (cf King \& Muldrew 2016) and spread over $\sim 4\pi$ steradians.

An observer viewing one of these systems from a location within the beam solid angle $4\pi b$ (where $b \ll 1$) would measure a very high specific intensity, and if they
assumed isotropic emission, would assign an apparent luminosity  $L_{\rm sph} = (1/b)L \gg L$, where $L$ is the true total power output of the source. In this case the deduction of unusually large black hole masses from the assumption that $L_{\rm Edd} > L_{\rm sph}$ is unjustified, as it implies a minimum apparent mass $M_{\rm sph}$ greater than the true minimum value by a factor $1/b \gg 1$. An observer viewing a system of this type from outside the beam (i.e. by far the more common case) would assign an apparent luminosity $\sim L_{\rm Edd}$ because of the much lower specific intensity. So even though even though the system's apparent luminosity is very high when viewed along the beam, this substantial near--isotropic luminosity seen in all other directions must in all cases characterize the source's effect on its surroundings, as for example seen in emission line data, which are therefore the same whether one views the beamed emission or not.

Very recently, X--ray polarimetric observations by Veledina et al. (2023) have effectively settled two decades of controversy over the cause of  ULX behaviour, concluding 
`unambiguously' that
much of the accretion energy produced by the stellar--mass X--ray binary
 Cyg X-3 does follow two narrow 
channels (in that particular system pointing away from the observer). 

This agrees closely with the picture suggested by King et al. (2001) (see Lasota \& King 2023 for a discussion) which implies that ULX behaviour is simply a phase of standard X--ray binary evolution characterized by very high mass transfer rates. This phase occurs whenever a donor star which is more massive than the accretor fills its Roche lobe. Then mass transfer decreases the binary separation and amplifies the mass transfer, until this reaches the (very high) limiting rate set by the donor star's thermal timescale.
I will show later in the paper that high--redshift supermassive black holes may 
at some point have to deal with similarly super--Eddington mass supply rates, producing effects like those seen in ULXs, and consequent problems for deducing masses. 

For ULXs,
King (2009) showed that quite generally,
\be
b \simeq \frac{73}{{\dot m}^2}.
\label{beam}
\ee
 This formula agrees with simple physical reasoning about the open funnel geometry (King, 2009), since conditions at large disc radii are set by $\dot M_{\rm tr}$, while if $\dot m > 1$, conditions at small disc radii converge to values set by the Eddington rate $\dot M_{\rm Edd}$.  Assuming normal stellar--mass black hole and neutron star accretors,
equation (\ref{beam}) gives acceptable fits to all observations of ULXs (see King et al., 2023 for a review).

\section{Mass Estimates and Outflows}

The results summarized in the last Section, and the scale--free nature of black hole accretion, suggest that it may be interesting to consider what observations of high--$z$ quasars would tell us if the sources are assumed beamed like ULXs, as a result of being fed mass at super--Eddington rates, rather than radiating isotropically. Since black--hole accretion is scale--free we can assume that  (\ref{beam}) holds for such ULX--like quasars. Note that the argument that Haiman \& Cen (2002) made against quasar beaming does not apply to the picture I consider here -- they considered a quasar model beamed towards the observer, but which they explicitly assumed had a low total luminosity. In constrast we saw in the last Section that ULX--type emission is always accompanied by a substantial near--isotropic component, which would evidently produce the same emission--line properties as an unbeamed quasar.

The first consequence is that for ULX--like quasars the luminosity alone no longer requires unusually high black hole masses in order to avoid conflict with the Eddington limit, just as one had the freedom to consider stellar--mass ULX models rather than the original IMBH proposal for explaining the original ULX systems.

But the deduction of large masses in high--redshift quasars generally comes from the use of virial mass indicators rather than the quasar luminosities (cf Willott, McLure \& Jarvis, 2003). Virial indicators use the substantial observed velocity widths $\Delta V$ of emission lines such as H$\beta$ and Mg~II to deduce the SMBH mass by assuming the the gas is in bound orbits in the AGN's broad--line region (BLR). At low redshift, reverberation mapping of AGN gives the response time of the BLR to central AGN luminosity changes. This fixes the BLR size $R$, and allows a black--hole mass estimate
\be
M \sim \frac{R(\Delta V)^2}{G}.
\ee

At high redshift the much larger expected size of the BLR (several light years) makes reverberation mapping difficult, so the SMBH mass estimates proceed by extrapolating the correlation between continuum luminosity and BLR size observed for low--redshift systems. But a serious problem in using this virial method to estimate masses for the hypothesised ULX--like quasars is that the spectra of stellar--mass ULXs are dominated by strong and fast outflows. The velocity widths then do not correspond to motions of line--emitting material in bound orbits about the SMBH, but radial winds with the escape speed from the black hole's vicinity.

In stellar--mass ULXs these winds remove most of the super--Eddington mass supply rate $\dot M$, and have observed velocity widths 
$> 0.2c$ (e.g. Walton et al., 2016). This agrees with the values expected for 
momentum--driven flows, where the gas velocity $v$ is given by equating outflow and radiation momenta as
\be
\dot Mv \simeq \frac{L_{\rm Edd}}{c},
\ee
which implies a radial outflow speed
\be
v \simeq \frac{\eta }{\dot m}c,
\ee
where $\eta \sim 0.1$ is the accretion efficiency. As expected, this gives no direct information about the black hole mass, as the problem is scale--free. 

Momentum--driven `UFO'  winds like this, characterized by P Cygni profiles in their X--ray spectra, 
are very widely observed in bright AGN (e.g. Tombesi et al., 2010a, b and many subsequent papers). Since the velocity width is no longer given by orbital motion of bound gas in the BLR we cannot use it to deduce the mass of the central supermassive black hole if the AGN are `ULX--like'.\footnote{
Physically, these UFO winds correspond to cases where the reverse shock produced where the outflow collides with the host galaxy interstellar gas is efficiently cooled, typically by the inverse Compton effect of the AGN radiation field (see King \& Pounds, 2015 and King, 2023 for reviews). This prevents the thermal expansion of the shocked gas driving much stronger and faster energy--driven outflows. But for a sufficiently large SMBH mass, Compton shock cooling becomes inefficient, the outflows become energy--driven, and now drive off all of the accreting gas. This fixes the SMBH mass, and is the likely origin of the $M - \sigma$ relation.}

\section{Masses and Distances}

The quadratic dependence of the beaming factor $b$ (see eqn \ref{beam}) on $\dot m = \dot M/\dot M_{\rm Edd}$ means that deducing limits on accretor masses in disc--wind beamed systems is more complex than for isotropic emission.

Both $\dot M_{\rm Edd}$ and $L_{\rm Edd}$ are proportional to the accretor mass $M$, so  for a given mass supply rate $\dot M$ we have
\be
b \propto \left(\frac{M}{\dot M}\right)^2
\ee
giving the apparent (assumed isotropic) luminosity 
\be
L_{\rm sph} \simeq \frac{L_{\rm Edd}}{b} \propto \frac{\dot M^2}{M},
\ee
i.e. the tighter beaming for a lower--mass accretor outweighs its lower Eddington luminosity when we predict its apparent luminosity.

Then the standard procedure of deducing an accretor mass $M_{\rm sph}$ by arguing that the {\it apparent} luminosity
$L_{\rm sph}$ cannot exceed the Eddington limit for the corresponding mass $M_{\rm sph}$ results in an estimate
\be
M_{\rm sph} \propto \frac{\dot M^2}{M}
\label{M}
\ee
for the minimum accretor mass.
Thus in an ensemble of beamed sources with similar super--Eddington mass supply rates $\dot M$, the sources with the highest apparent luminosities actually have the {\it lowest} masses because their beaming is quadratically tighter.

But beaming means that many of these sources will not be observable. In the simple case where redshifts are modest, to lie in the beam of one such source one has to search a space volume $\sim 1/Nb$, where $N$ is the spatial number density of the sources. Then the nearest detectable beamed source is at a distance
\be
D \sim \left(\frac{3}{4\pi Nb}\right)^{1/3} \propto \left(\frac {\dot M}{NM}\right)^{2/3}.
\label{D}
\ee
We see that in observations of a collection of accreting sources with similar mass supply rates $\dot M$, beamed according to (\ref{beam}), 

\bigskip
\noindent
{\it the systems with the highest apparent luminosities and apparent masses are in reality the most distant, i.e. the most redshifted, of those sources with the lowest black hole masses\footnote{In the stellar--mass case, King (2009) used these arguments to point out that some extragalactic sources usually assumed to be AGN might actually be ULXs (`pseudoblazars'), distinguishable from real AGN only by the fact that they are not constrained to lie in the dynamical centres of their host galaxies.}.}
\bigskip

If these same beamed sources were instead assumed to radiate isotropically,
we would deduce that the highest redshift sources were intrinsically the brightest, and because of the Eddington limit, must have very high black hole masses. This is exactly the result one gets using the straightforward interpretation of the data on distant quasars. 


These very simple arguments already suggest that ULX--type beaming may have interesting implications for the discussion of the masses of high--redshift QSOs. But it is clear that taking these ideas further requires careful consideration of the evolution of number density $N$ with redshift. 

\section{Discussion}

The arguments given in this paper suggest that high--redshift QSO central black holes
may have relatively modest masses if in reality they are fed at potentially strongly super--Eddington rates. 

This is already known to hold for stellar--mass ULX systems. 
Extending this picture to high--$z$ QSOs is reasonable, since at the relevant redshifts it is likely that some black hole masses may not have reached large values, and for example may lie below the $M - \sigma$ value
\be
M_{\sigma} \simeq 3.2\times 10^8\msun\sigma_{200}^4
\label{msig}
\ee
(e.g. Kormendy \& Ho, 2013)
corresponding to the velocity dispersion $\sigma = 200\sigma_{200}\,{\rm km\, s^{-1}}$ of their host galaxy bulges.
If the galaxy is strongly disturbed (e.g. by a merger), mass would be supplied at a dynamical rate
\be
\dot M_{\rm dyn} \lesssim \frac{f_g\sigma^3}{G},
\ee
where $f_g \simeq 0.16$ is the cosmological gas fraction,
giving an Eddington factor
\be
\dot m \sim \frac{\dot M_{\rm dyn}}{\dot M_{\rm Edd}} =
\frac{54}{M_8^{1/4}}\frac{M_{\sigma}}{M}
\ee
(cf King \& Pounds, 2015; King, 2023). From (\ref{beam}) this implies typical beaming fractions $ b \lesssim  10^{-2}$, with corresponding reductions in the deduced true SMBH mass $\sim bM_{\rm sph}$.

Evidently the condition for ULX--type 
disc--wind beaming of the accretion luminosity is readily satisfied in many high--redshift galaxies. Given the scale--free property of black hole accretion it would be surprising if Nature made use of this effect for stellar--mass black holes only.

Future observations should provide direct tests of these ideas. For stellar--mass ULXs 
X--ray polarimetry revealed the beaming. For AGN systems at nonzero redshift the relevant photon energy is softer, and it is unclear whether a test like this would be possible, and if so, conclusive. The recent discovery of the very bright QSO system J0529–4351 (Wolf et al., 2024) may allow direct measurement of its large expected BLR size, and so test the extrapolations used in current methods for estimating black hole masses in high--$z$ QSOs. Theoretically, calculations allowing one to discriminate between emission lines from outflows and from orbiting gas would be very useful.

\section*{DATA AVAILABILITY}
No new data were generated or analysed in support of this research.

\section*{ACKNOWLEDGMENTS}
I thank Jean--Pierre Lasota for many discussions of ULX behaviour, and the referee for thoughtful comments which improved the paper.

{}
\end{document}